%% file: main.tex
\documentclass[final,2p,times,twocolumn]{elsarticle}
\usepackage[utf8]{inputenc}

\usepackage{natbib}
\usepackage[font=normalsize]{caption}
\usepackage{graphicx}
\usepackage{amsmath, bm, amsfonts}
\usepackage{stmaryrd}
\usepackage{multirow, adjustbox, chngpage}
\usepackage{bigstrut}
\usepackage{fancybox}
\usepackage[flushleft]{threeparttable}
\usepackage{color}
\usepackage{balance}
\usepackage[top=1in, textwidth=19cm]{geometry}
\journal{Journal of Systems and Software}

\newcommand{\NAME}{SMAD}
\newcommand{\RQone}{Does \NAME{} outperform standalone detection tools?}
\newcommand{\RQtwo}{Does \NAME{} outperform other ensemble methods?}

\DeclareMathOperator*{\maxx}{max}
\DeclareMathOperator*{\argmax}{argmax}

\newcommand{\Answer}[1]{\begin{center}%
    \noindent\thicklines\setlength{\fboxsep}{8pt}%
    \cornersize{0.2}\Ovalbox{\begin{minipage}{8.5cm}%
    \textit{#1}\end{minipage}} \end{center}}

\begin{document}
\begin{frontmatter}
\title{A Machine-learning Based Ensemble Method For Anti-patterns Detection}

\author[poly]{Antoine Barbez}
\ead{antoine.barbez@polymtl.ca}

\author[poly]{Foutse Khomh}
\ead{foutse.khomh@polymtl.ca}

\author[concordia]{Yann-Gaël Guéhéneuc}
\ead{yann-gael.gueheneuc@concordia.ca}

\address[poly]{Polytechnique Montreal}
\address[concordia]{Concordia University}

\begin{abstract}
Anti-patterns are poor solutions to recurring design problems. Several empirical studies have highlighted their negative impact on program comprehension, maintainability, as well as fault-proneness. A variety of detection approaches have been proposed to identify their occurrences in source code. However, these approaches can identify only a subset of the occurrences and report large numbers of false positives and misses. Furthermore, a low agreement is generally observed among different approaches. Recent studies have shown the potential of machine-learning models to improve this situation. However, such algorithms require large sets of manually-produced training-data, which often limits their application in practice. 

In this paper, we present SMAD (SMart Aggregation of Anti-patterns Detectors), a machine-learning based ensemble method to aggregate various anti-patterns detection approaches on the basis of their internal detection rules. Thus, our method uses several detection tools to produce an improved prediction from a reasonable number of training examples. We implemented SMAD for the detection of two well known anti-patterns: God Class and Feature Envy. With the results of our experiments conducted on eight java projects, we show that: (1) our method clearly improves the so aggregated tools; (2) SMAD significantly outperforms other ensemble methods.
\end{abstract}

\begin{keyword}
Software Quality, Anti-patterns, Machine Learning, Ensemble Methods
\end{keyword}

\end{frontmatter}

\input{01Introduction}
\input{02RelatedWork}
\input{03SMAD}
\input{04Methodology}
\input{05Study}

\input{06Threats}
\input{07Conclusion}
\section*{Acknowledgement}
This work is partly supported by the The Natural Sciences and Engineering Research Council of Canada (NSERC) and the Canada Research Chair on Patterns in Mixed-language Systems.
\section{References}
\balance
\bibliographystyle{plain}
\bibliography{references}
\end{document}

%% file: 01Introduction.tex
\section{Introduction}
Source code refactoring, which consists in improving the internal structure of the code while keeping its external behaviour unchanged, represents a substantial portion of software maintenance activities.
To help practitioners in identifying where -- and what kind of -- refactoring should be applied in software systems, the concept of \textit{design smells} has been introduced and defined by Fowler~\cite{Fowler1999} as symptoms of poor solutions to recurring design problems. These symptoms, also called \textit{anti-patterns}, are typically introduced in object-oriented systems when developers implement sub-optimal design solutions due to lack of knowledge and--or time constraints. For example, the God Class anti-pattern refers to the situation in which a class grows rapidly by the addition of new functionalities, when developers break the principle of single responsibility. Prior empirical studies highlighted the negative impact of anti-patterns on a variety of quality characteristics, such as program comprehension~\cite{abbes2011empirical}, maintainability~\cite{yamashita2013exploring}, and correctness (increase of fault-proneness)~\cite{khomh2012exploratory}. Thus, it is important to identify their occurrences in systems and apply refactoring operations to remove them.

Several approaches have been proposed to detect the occurrences of anti-patterns in source code. Most of these approaches attempt to identify bad motifs in models of source code using manually-defined heuristics that rely on some metrics (e.g., cyclomatic complexity). For example, Moha et al.~\cite{Moha10-TSE-DECOR} proposed a domain-specific language to describe and generate detection algorithms for anti-patterns using structural and lexical metrics while Palomba et al.~\cite{PalombaBPOLP13,Palomba15} proposed a rule-based approach to detect anti-patterns from change history information.

Even though these approaches have shown acceptable performances, none of them can claim high accuracy on any systems and for any anti-patterns. Besides, each approach relies on its own definitions of anti-patterns and only focuses on specific aspects of systems. Thus, we observe a complementarity among the different approaches, especially when they rely on orthogonal sources of information (e.g., structural vs. historical) \cite{fontana2012automatic, PalombaBPOLP13}.

Consequently, we propose SMAD (SMart Aggregation of Anti-patterns Detectors), a machine-learning based ensemble method to combine various anti-patterns detection approaches in order to achieve better performances. Hence, our approach aims at reducing software maintenance costs by helping practitioners in identifying more accurately the code components to be refactored. Concretely, for each approach to be aggregated, we identify a set of core metrics, i.e., metrics that reflect the internal detection rule of the approach. We then use the core metrics as input features of a neural-network classifier. To the best of our knowledge, we are the first to propose an ensemble method in the context of anti-patterns detection.

Recently, machine-learning models have been shown efficient in a variety of domains, such as speech recognition~\cite{graves2013speech} or image processing~\cite{krizhevsky2012imagenet}. Several machine-learning based approaches have been proposed to detect anti-patterns. However, these approaches failed to surpass clearly conventional techniques. On the one hand, learning high-level features of systems requires complex machine-learning models, such as deep-neural-networks. On the other hand, these complex models require substantial amounts of manually-produced training data, which is hardly available and time consuming to produce for anti-patterns.

On the contrary our method relies on existing approaches which allows our model to take as input a low number of high-level key features (i.e.,  the core metrics). As a consequence, our method can benefit from a simple machine-learning classifier that requires a reasonable number of training examples. We implemented the proposed ensemble method to detect two well known anti-patterns: God Class and Feature Envy. To conduct our experiments, we created an oracle containing occurrences of the studied anti-patterns in eight Java systems. 

This paper thus makes the following contributions: (1) a manually-produced oracle reporting the occurrences of God Class and Feature Envy in eight Java software systems; (2) a machine learning-based ensemble method to aggregate existing anti-patterns detection approaches.

The remainder of this paper is organized as follows. Section~\ref{section: related work} defines the two anti-patterns considered in this study and discusses the related work. Section~\ref{section: smad} describes our approach \NAME{}. Section~\ref{section: methodology} presents our data as well as preliminary considerations for our study. Section~\ref{section: evaluation performances} reports the results of our study aiming to evaluate the performances of our method as well as to compare it with other approaches. Section~\ref{section: threats} discusses the threats that could affect the validity of our results. Finally, Section~\ref{section: conclusion} concludes and discusses future work.

%% file: 02RelatedWork.tex
\section{Background and Related Work}
\label{section: related work}
This section defines the anti-patterns considered in this study and discusses prior detection approaches proposed in literature, as well as ensemble methods.
\subsection{Definitions}
\subsubsection{God Class}
A God Class or Blob, is a class that tends to centralize most of the system's intelligence, and implements a high number of responsibilities.
It is characterized by the presence of a large number of attributes, methods and dependencies with data classes (i.e., classes only used to store data in the form of attributes that can be accessed via getters and setters). Thus, assigning much of the work to a single class, delegating only minor operations to other small classes causes a negative impact on program comprehension \cite{abbes2011empirical} and reusability. The alternative refactoring operation commonly applied to remove this anti-pattern is called Extract Class Refactoring and consists in splitting the affected God Class into several more cohesive smaller classes \cite{Fowler1999}.

\subsubsection{Feature Envy}
A method that is more interested in the data of another class (the envied class) than that of the class it is actually in. This anti-pattern represents a symptom of the method's misplacement, and is characterized by a lot of accesses to foreign attributes and methods. The main consequences are an increase of coupling and a reduction of cohesion, because the affected method often implements responsibilities more related to the envied class with respect to the methods of its own class. This anti-pattern is commonly removed using Move Method Refactoring, which consists in moving all or parts of the affected method to the envied class \cite{Fowler1999}.

\subsection{Rule Based Approaches}
The first attempts to detect components affected by anti-patterns have focused on the definition of rule-based approaches which rely on some metrics to capture deviations from good object-oriented design principles.
First, Marinescu \cite{marinescu2004detection} presented \textit{detection strategy}, a metric-based mechanism for analyzing source code models and detect design fragments. They illustrate this methodology step by step by defining the detection strategy for God Class. Later, Lanza and Marinescu \cite{lanza2007object} formulated the detection strategies for 11 anti-patterns by designing a set of metrics, along with thresholds, for each anti-pattern. These metric--threshold pairs are then logically combined using AND/OR operators to create the final detection rule. These heuristics have been implemented inside Eclipse plug-ins such as \textit{InCode} \cite{marinescu2010incode}.

Similar to the approach described above, Moha et al.~\cite{Moha10-TSE-DECOR} proposed DECOR (DEtection and CORrection of Design Flaws) which relies on a systematic analysis of the definitions of code and design smells in the literature. They propose templates and a grammar to encode these smells and generate detection algorithms automatically. They applied their approach to four design anti-patterns (God Class, Functional  Decomposition, Spaghetti Code, and Swiss Army Knife) and their 15 underlying code smells. Their detection approach takes the form of a ``Rule Card'' that encodes the formal definition of design anti-patterns and code smells. In this context, the identification of components affected by a God Class is based on both structural and lexical information.

Other approaches rely on the identification of refactoring opportunities to detect anti-patterns. Based on this consideration, instances of a given anti-pattern can be detected in a system by looking at the opportunities to apply the corresponding refactoring operation. In this context, Fokaefs et al.~\cite{fokaefs2012identification} proposed an approach to detect God Classes in a system by suggesting a set of Extract Class Refactoring operations. This set of refactoring opportunities is generated in two main steps. First, they identify cohesive clusters of entities (i.e., attributes and methods) in each class of the system, that could then be extracted as separate classes. To do so, the Jaccard distance is computed among each class members (i.e., entities). The Jaccard distance between two entities $e_{i}$ and $e_{j}$ measures the dissimilarity between their respective \textit{``entity sets''} $S_{i}$ and $S_{j}$ and is computed as follows:
\begin{equation}
\label{jaccard entity to entity}
dist(e_{i}, e_{j}) = 1 - \frac{|S_{i} \cap S_{j}|}{|S_{i} \cup S_{j}|}
\end{equation}
For a method, the \textit{``entity set''} contains the entities accessed by the method, and for an attribute, it contains the methods accessing this attribute. Then, cohesive groups of entities are identified using a hierarchical agglomerative algorithm on the information previously generated. In the second step, the potential classes to be extracted are filtered using a set of rules, to ensure that the behavior of the original program is preserved.
Later, this approach has been implemented as an Eclipse plug-in called \textit{JDeodorant} \cite{fokaefs2011jdeodorant}.

Similarly, methods that can potentially be moved to another class are presented to the software engineer as potential Feature Envy methods. In this context, Tsantalis and Chatzigeorgiou~\cite{tsantalis2009identification} proposed an approach for automatic suggestions of Move Method Refactoring. First, for each method $m$ in the system, a set of candidate target classes $T$ is created by examining the entities that are accessed in the body of $m$. Second, $T$ is sorted according to two criteria: (1) the number of entities that $m$ accesses from each target class of $T$ in descending order and; (2) the Jaccard distance from $m$ to each target class in ascending order if $m$ accesses an equal number of entities from two or more classes. In this context, the Jaccard distance between an entity $e$ and a class $C$ is computed as follows:
\begin{equation}
\label{jaccard entity to class}
dist(e, C) = 1 - \frac{|S_{e} \cap S_{C}|}{|S_{e} \cup S_{C}|} \quad \textrm{where} \quad S_{c} = \bigcup_{e \in C} \{e\}
\end{equation}
With $S_{e}$ the entity set of a method defined in Equation~\ref{jaccard entity to entity}.
Third, $T$ is filtered under the condition that $m$ must modify at least one data structure in the target class. Fourth, they suggest to move $m$ to the first target class in $T$ that satisfies a set of preconditions related to compilation, behavior, and quality. This algorithm is also implemented in the Eclipse plug-in \textit{JDeodorant} \cite{fokaefs2007jdeodorant}.

Anti-patterns can also impact how source code entities evolve with one another over time, when changes are applied to the system. Based on such consideration, Palomba et al.~\cite{PalombaBPOLP13,Palomba15} proposed HIST (Historical Information for Smell deTection), an approach to detect anti-patterns using historical information derived from version control systems (e.g., Git, SVN). They applied their approach to the detection of five anti-patterns: Divergent Change, Shotgun Surgery, Parallel Inheritance, God Class and Feature Envy. The detection process followed by HIST consists of two steps. First, historical information is extracted from versioning systems using a component called the \textit{change history extractor} which outputs the sequence of changes applied to source code entities (i.e., classes or methods) through the history of the system. Second, a set of rules is applied to this so produced sequence to identify occurrences of anti-patterns. For instance, Feature Envy methods are identified as those ``\textit{involved in commits with methods of another class of the system $\beta$ \% more than in commits with methods of their class}''. The value of $\beta$ being set to 80\% after parameters calibration.

\subsection{Machine Learning Based Approaches}
Kreimer~\cite{kreimer2005adaptive} proposed the use of decision trees to identify occurrences of God Class and Long Method. Their model relies on the number of fields, number of methods, and number of statements as decision criteria for God Class detection and have been evaluated on two small systems (IYC and WEKA). This observation has been confirmed 10 years later by Amorim et al.~\cite{amorim2015experience} who extended this approach to 12 anti-patterns.

Khomh et al.~\cite{khomh2009bayesian, khomh2011bdtex} presented BDTEX (Bayesian Detection Expert), a metric based approach to build Bayesian Belief Networks from the definitions of anti-patterns. This approach has been validated on three different anti-patterns (God Class, Functional Decomposition, and Spaghetti Code) and provides a probability that a given entity is affected instead of a boolean value like other approaches. Following, Vaucher et al.~\cite{vaucher2009tracking} relied on Bayesian Belief Networks to track the evolution of the``godliness'' of a class and thus, distinguishing real God Classes from those that are so by design.

Maiga et al.~\cite{maiga2012smurf,maiga2012support} introduced SVMDetect, an approach based on Support Vector Machines to detect four well known anti-patterns: God Class, Functional Decomposition, Spaghetti code, and Swiss Army Knife. The input vector fed into their classifier for God Class detection is composed of 60 structural metrics computed using the PADL meta-model~\cite{gueheneuc2005ptidej}.

Fontana et al.~\cite{fontana2016comparing} performed the largest experiment on the effectiveness of machine learning algorithms for smell detection. They conducted a study where 16 different machine learning algorithms were implemented (along with their boosting variant) for the detection of four smells (Data Class, God Class, Feature Envy, and Long Method) on 74 software systems belonging to the \texttt{Qualitas Corpus} dataset \cite{tempero2010qualitas}. The experiments have been conducted using a set of independent metrics related to class, method, package and project level as input information and the datasets used for training and evaluation have been filtered using an under-sampling technique (i.e., instances have been removed from the original dataset) to avoid the poor performances commonly reported from machine learning models on imbalanced datasets. Their study concluded that the algorithm that performed the best for both God Class and Feature Envy was the J48 decision tree algorithm with an F-measure close to 100\%. However, Di Nucci et al.~\cite{di2018detecting} replicated their study and highlighted many limitations. In particular, the way the datasets used in this study have been constructed is strongly discussed and the performances achieved after replication were far from those originally reported.

More recently, Liu et al.~\cite{liu2018deep} proposed a deep learning based approach to detect Feature Envy. Their approach relies on both structural and lexical information. On one side, the names of the method, the enclosing class (i.e., where the method is implemented) and the envied class are fed into convolutional layers. On the other side, the distance presented in Equation~\ref{jaccard entity to class} is computed for both the enclosing class ($dist(m, ec)$) and the target class ($dist(m, tc)$), and values are fed into other convolutional layers. Then the output of both sides is fed into fully-connected layers to perform classification. To train and evaluate their model, they use an approach similar to Moghadam and Ó Cinnéide~\cite{moghadam2012automated} where labeled samples are automatically generated from open-source applications by the injection of affected methods. These methods assumed to be correctly placed in the original systems are extracted and moved into random classes to produce artificial Feature Envy occurrences (i.e., misplaced methods).

\subsection{Ensemble Methods}
Ensemble methods aim at aggregating multiple classifiers in order to improve the classification performances. Ensemble methods are commonly used in the literature as Boosting techniques \cite{rokach2010ensemble}, i.e., a function is applied to the output of various machine-learning based classifiers trained independently. We found no existing ensemble method proposed for anti-patterns detection, as most of the existing detection approaches rely on manually defined rules. However, two ensemble methods proposed in the context of Bug prediction can be applied to our problem. These methods aim at aggregating the output of multiple machine-learning based classifiers (e.g., Support Vector Machine, Multi-layer Perceptron, Decision Tree) trained to predict the bug-proneness of software components.

First, Liu et al.~\cite{liu2010evolutionary} proposed Validation and Voting, which consists in considering a majority vote over the outputs of the classifiers. Second, Di Nucci et al.~\cite{di2017dynamic} proposed ASCI (Adaptive Selection of Classifiers in bug predIction), an adaptive method which uses a decision tree algorithm to dynamically predict which classifier is the best for each code component to be classified. The workflow of ASCI is organised as follows: Given a training set $T$ of instances (i.e., code components) and their associated labels (i.e., buggy or not buggy), each classifier is experimented against $T$. Then a new training set $T^{*}$ is created by labelling each input instance with the information about the classifier which correctly identified its bug-proneness. Then a decision tree is trained on $T^{*}$ to predict the best classifier for each input instance.    

%% file: 03SMAD.tex
\section{SMAD: SMart Aggregation of Anti-patterns Detectors}
\label{section: smad}
\subsection{Problem Definition}
Let us consider $D$ anti-patterns detection tools $d_{1},d_{2}, ..., d_{D}$ performing a boolean prediction over the entities (i.e., classes or methods) of a software system based on some internal detection rule. We refer to as $d_{i}(e) \in \{True, False\}$ the boolean prediction of the $i^{th}$ detection tool on an entity $e$.

We want to combine these approaches to maximize the Matthew's Correlation Coefficient (MCC) (cf. Equation~\ref{mcc}) of the so-produced ``merged'' prediction over the entities of the studied system. Thus, we want to find a function of the $D$ approaches that outputs an improved prediction on any software system $S$. Which can be expressed as:

\[f(d_{1},d_{2}, ..., d_{D}, e) \in \{True, False\}\] 
and 
\[MCC(f(d_{1},d_{2}, ..., d_{D}, S)) \geq \maxx_{i} MCC(d_{i}(S)), \quad \forall S\]

\subsection{Overview}
\label{subsection: overview}
The proposed method \NAME{} allows to combine various detection approaches on the basis of their internal detection rule instead of their output. Indeed, the internal detection process of any rule based approach relies on some structural or historical metrics (i.e., core-metrics) that can be computed for any entity to be classified. Thus, the key idea behind \NAME{} is to compute the core-metrics of various approaches for each input entity and use these metrics to feed a machine-learning based classifier. First, for each anti-pattern considered in this study, we selected three state-of-the-art detection tools to be aggregated. These tools respectively rely on:

\begin{itemize}
\item \textit{Rule Cards:} Affected entities are identified using a combination of source-code metrics designed to reflect the formal definition of the anti-patterns. For this category, we selected DECOR \cite{Moha10-TSE-DECOR} for God Class and InCode \cite{marinescu2010incode} for Feature Envy detection.  

\item \textit{Historical Information:} Affected entities are identified via an analysis of change history information derived from versioning systems. For this category, we used HIST \cite{PalombaBPOLP13,Palomba15} for both God Class and Feature Envy detection.

\item \textit{Refactoring Opportunities:} Anti-patterns are detected by identifying the opportunities to apply their corresponding refactoring operations. For this category, we used the refactoring operations Extract Class \cite{fokaefs2012identification} and Move Method \cite{tsantalis2009identification} provided by JDeodorant, respectively for God Class and Feature Envy detection.
\end{itemize}

\begin{figure}[htb]
\centering
\includegraphics[width=9.2cm]{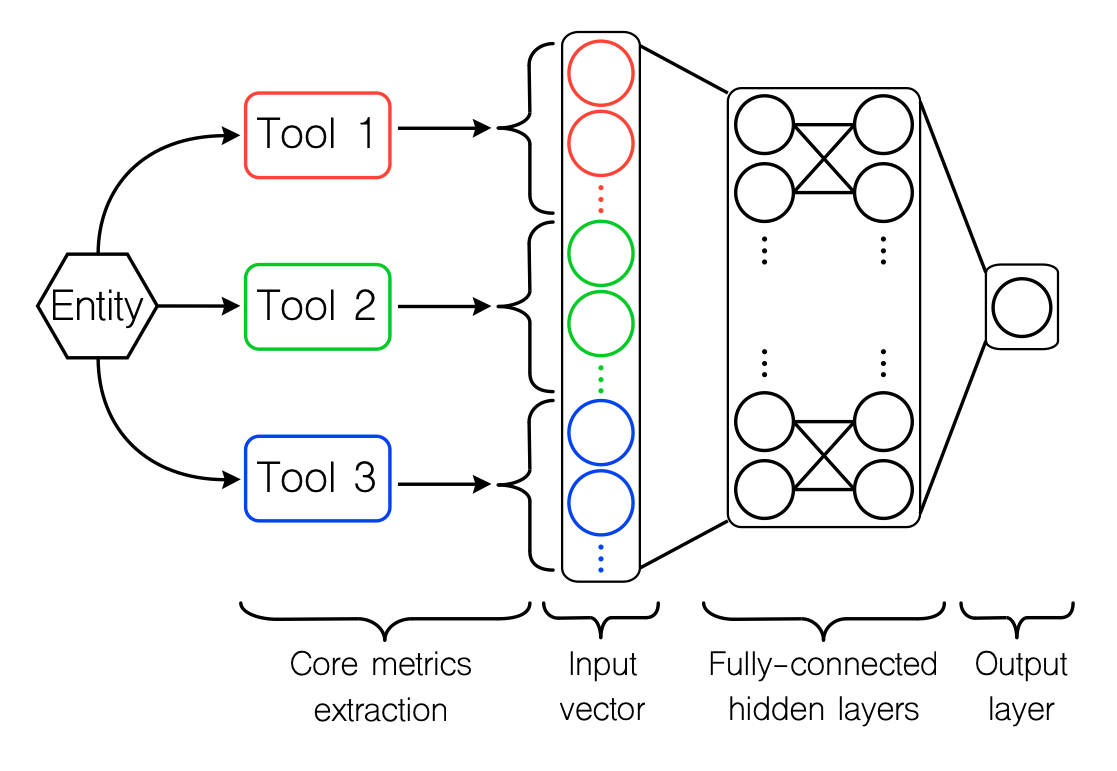}
\caption{Overview of SMAD Detection Process}
\label{fig: smad}
\end{figure}

Selecting tools that are based on different strategies allows us to expect a high complementarity of the aggregated approaches and thus, maximize the expected performances of our method. Then, we selected the core metrics, i.e., metrics that reflect best the internal decision process of each tool, as input features for our model. The classifier used by \NAME{} to predict whether an entity is an anti-pattern or not is a Multi-layer Perceptron (MLP), i.e., a fully-connected feed-forward neural-network. This model is composed of \textit{tanh} dense hidden layers connected to a \textit{sigmoid} output layer. Fig.~\ref{fig: smad} overviews our approach.

\subsection{Input}
\label{subsection: input}
\subsubsection{Metrics for God Class Detection}

For God Class detection, we extract \textbf{six} core metrics from the three detection tools considered in this study. These metrics are computed for each class of a system.

\paragraph{DECOR:}
The internal detection rule relies on the definition of four code and design smells and can be expressed as: ``(is associated to many \textit{DataClass}) AND is a (\textit{ControllerClass} OR \textit{LargeClass} OR \textit{LowCohesionClass})''. These smells are defined using structural and lexical metrics along with some thresholds: (1) \textit{DataClass} relies on the number of accessors, (2) \textit{ControllerClass} relies on lexical properties, (3) \textit{LargeClass} relies on the  sum of the NMD and NAD metrics (Number of Methods Declared + Number of Attributes Declared), and (4) \textit{LowCohesionClass} relies on the LCOM metric (Lack of Cohesion in Methods) \cite{briand1998unified}. Thus, we extracted four core metrics from DECOR internal detection rule for God Class:
\begin{itemize}
\item  Number of associated \textit{DataClass}
\item $\llbracket$\textit{ControllerClass}$\rrbracket$
\item nmd\_nad
\item lcom 
\end{itemize}

\noindent with $\llbracket \rrbracket$ a boolean variant of the kronecker delta defined in Equation~\ref{kronecker}, nmd\_nad and lcom are the uppercase metrics divided by their respective threshold. The value of these thresholds depends on the systems characteristics and can be computed through the Ptidej API~\footnote{https://github.com/ptidejteam/v5.2}.

\paragraph{HIST:}
God Classes are identified as: ``classes modified (in any way) in more than $\alpha$\% of commits involving at least another class''. Thus, we extracted one core metric from HIST internal detection rule for God Class:
\begin{itemize}
\item  Ratio between the number of commits in which the considered class has been modified along with other classes and the number of commits involving other classes of the system.
\end{itemize}

\paragraph{JDeodorant:}
God Classes are classes from which a \textit{concept} (i.e., a subclass) can be extracted. Formally, a concept is defined as: ``a distinct entity or abstraction for which a single class provides a description and/or a set of attributes and methods that contribute together to the same task''. We define our core metric for JDeodorant as:
\begin{itemize}
\item  Number of \textit{concepts} that can be extracted from the considered class.
\end{itemize}

\subsubsection{Metrics for Feature Envy Detection}
\label{Paragragh: metrics feature envy}

A Feature Envy is characterized by two source code components: a method (i.e., the envious method) and a class (i.e., the envied class). Hence, in a given system, the number of potential entities that must be investigated is equal to $n_{m}\times(n_{c}-1)$ with $n_{c}$ and $n_{m}$ respectively the numbers of classes and methods in the system. To reduce this number, we filter the studied system at both class and method level, similarly to Tsantalis and Chatzigeorgiou~\cite{tsantalis2009identification}. First, we consider as potential envious methods only non-static and non-accessor methods. Then, for each of the remaining methods, we consider as potential envied classes only classes that are accessed in some way in the body of the method. We extract \textbf{seven} core metrics from the three considered detection tools.

\paragraph{InCode:}
Methods are identified as being envious without information about the envied class. In this context, a method is declared affected if: (1) ``it uses directly more than a few attributes of other classes'' (ATFD $>$ FEW), (2) ``it uses far more attributes from other classes than its own'' (LAA $<$ ONE THIRD), and (3) ``the used “foreign” attributes belong to very few other classes'' (FDP $\leq$ FEW). We redefined the first two metrics to express information about the envied class, which led us to three core metrics:
\begin{itemize}
\item ATFD (Access To Foreign Data), i.e., number of attributes of the envied class accessed by the method.
\item LAA (Locality of Attribute Accesses), i.e., ratio between the number of accesses to attributes that belongs to the envied class vs.\ the enclosing class.
\item FDP (Foreign Data Providers), i.e., number of distinct foreign classes whose attributes are accessed by the method. 
\end{itemize}

\paragraph{HIST:}
Feature Envy methods are identified as: ``methods involved in commits with methods of another class of the system $\beta$\% more than in commits with methods of their class''. Thus, we extract one core metric from HIST internal detection rule for Feature Envy:
\begin{itemize}
\item Ratio between the number of commits involving the considered method and methods of the envied class vs. methods of the enclosing class. 
\end{itemize}

\paragraph{JDeodorant:}
For each method $m$ in the system, a set of candidate target classes $T$ is created and sorted based on: (1) the number of entities (methods or attributes) that $m$ accesses from each class of $T$ and (2) the Jaccard distance (cf. Equation~\ref{jaccard entity to class}) between $m$ and each target class. Then, JDeodorant suggests to move $m$ to the first target class that satisfies a set of preconditions related to compilation, behavior, and quality. We extracted three core-metrics, which can be expressed as:
\begin{itemize}
\item Ratio between the number of accesses to entities that belong to the envied class vs.\ enclosing class.
\item Ratio between the Jaccard distance from the method to the envied class vs.\ enclosing class.
\item Boolean value indicating whether the Move Method Refactoring operation has been proposed by JDeodorant or not.
\end{itemize}

\subsection{Replication Package}
To facilitate further evaluation and reuse of our work, all the data used in this study is publicly available\footnote{https://github.com/antoineBarbez/SMAD/}. Our replication package includes: (1) the oracle; (2) the source code of our model; (3) the scripts used to generate our data; and, (4) the implementation of our experiments. Furthermore, we created a component called \textit{repositoryMiner} to extract automatically all the necessary data to test or train our model on new systems.

%% file: 04Methodology.tex
\section{Methodology}
\label{section: methodology}
This section describes the data used to run our experiments, the metrics used to assess the performances of the different approaches investigated in this work, as well as the methodology used for training neural-networks to detect anti-patterns.
\subsection{Building a Reliable Oracle}
\label{Subsection: oracle}

To train and evaluate the performances of \NAME{}, as well as to compare it with competing classifiers, we needed an oracle reporting the occurrences of the two studied anti-patterns in a set of software systems. Unfortunately, we found no such large dataset in the literature. Indeed, creating a large oracle for anti-patterns is a painful process that requires weeks of manual source code analysis, which explains the unavailability of such data. Hence, we created our oracle following a different procedure for each of the two anti-patterns. 

For God Class, we found two sets of manually-detected occurrences in open-source Java systems, respectively from DECOR~\cite{Moha10-TSE-DECOR} and HIST~\cite{PalombaBPOLP13} replication packages\footnote{ http://www.rcost.unisannio.it/mdipenta/papers/ase2013/}\footnote{http://www.ptidej.net/tools/designsmells/materials/}. Thus, we created our oracle from these sets under two constraints: (1) the full history of the system must be available through Git or SVN, and (2) the occurrences reported must be relevant, i.e., we kept only the systems for which we agreed with the occurrences tagged. After filtering, over the 15 systems available in these replication packages, we retained eight to construct our oracle.  

For Feature Envy, most of the approaches proposed in the literature are evaluated on artificial occurrences, i.e., methods assumed to be correctly placed in the original systems, are then extracted and moved into random classes to produce Feature Envy occurrences (i.e., misplaced methods) \cite{moghadam2012automated, sales2013recommending, liu2018deep}. However, our approach partially relies on the history of code components. Therefore, such artificial occurrences are not usable because they have been willingly introduced in the considered systems' snapshot. Thus, we had to build manually our own oracle.

First, we formed a set of 779 candidate Feature Envy occurrences over the eight subject systems by merging the output of three detection tools (HIST, InCode, and JDeodorant), adjusting their detection thresholds to produce a number of candidate per system proportional to the systems sizes. Second, three different groups of people manually checked each candidate of this set: (1) the authors of this paper, (2) nine M.Sc.\ and Ph.D.\ students, and (3) two software engineers. We gave them access to the source code of the enclosing classes (where the methods were defined) and the potential envied classes. After analyzing each candidate, we asked respondents to report their confidence in the range [\textit{strongly\_approve}, \textit{weakly\_approve}, \textit{weakly\_disapprove}, \textit{strongly\_disapprove}]. To avoid any bias, none of the respondent was aware of the origin of each candidate. We made the final decision using a weighted vote over the reported answers. First we assigned the following weights to each confidence level:
\begin{center}
\resizebox{9.5cm}{!}{
\begin{tabular} {l l l l l l l l}
$strongly\_approve$ & $\rightarrow$ & $1.00$ && $weakly\_disapprove$ & $\rightarrow$ & $0.33$ \\
$weakly\_approve$ & $\rightarrow$ & $0.66$ && $strongly\_disapprove$ & $\rightarrow$ & $0.00$
\end{tabular}
}
\end{center}

Then, a candidate is considered as a Feature Envy if the mean weight of the three answers reported for this occurrence is greater than 0.5.

Table~\ref{Table: oracle} reports, for each system, the number of God Classes, the number of produced Feature Envy candidates, and the number of Feature Envy retained after manual-checking.

\begin{table}[htb]
\caption{Characteristics of the Oracle}
\label{Table: oracle}
\resizebox{9.2cm}{!}{
\begin{tabular} { l l l l}
\hline
System name &  \#God\_Class & \#Candidate\_FE& \#Feature\_Envy  \\ \hline
Android Opt Telephony&11&62&18 \\
Android Support&4&21&2 \\
Apache Ant&7&110&25 \\
Apache Lucene&4&42&4 \\
Apache Tomcat&5&173&57 \\
Apache Xerces&15&129&37 \\
ArgoUML&22&144&24 \\
Jedit&5&98&22 \\
\textbf{Total}&\textbf{73}&\textbf{779}&\textbf{189} \\ \hline
\end{tabular}
}
\end{table}

\subsection{Studied Systems}
\label{subsection: studied systems}
The context of our study consists of the eight open-source Java software systems presented in Table~\ref{Table: oracle}, which belong to various ecosystems. Two systems belong to the Android APIs\footnote{https://android.googlesource.com/}: Android Opt Telephony and Android Support. Four systems belong to the Apache Foundation\footnote{https://www.apache.org/}: Apache Ant, Apache Tomcat, Apache Lucene, and
Apache Xerces. Finally, one free UML design software: ArgoUML\footnote{http://argouml.tigris.org/} and one text editor: Jedit\footnote{http://www.jedit.org/} available under GNU General Public License\footnote{https://www.gnu.org/}. Without loss of generalizability, we chose to analyze only the directories that implement the core features of the systems and to ignore test directories. Table~\ref{Table: systems-hand} reports for each system, the Git identification (SHA) of the considered snapshot, its age (i.e., number of commit) and its size (i.e., number of class).

\begin{table}[htb]
\caption{Characteristics of the Studied Systems}
\label{Table: systems-hand}
\resizebox{9.2cm}{!}{
\begin{tabular} { l l l l l}
\hline
System name & Snapshot & Directory &\#Commit & \#Class \\ \hline
Android Opt Telephony&c241cad&src/java/&98&192 \\
Android Support&38fc0cf&v4/&195&109 \\
Apache Ant&e7734de&src/main/&6397&694\\
Apache Lucene&39f6dc1&src/java/&429&155\\
Apache Tomcat&398ca7ee&java/org/&3289&925\\
Apache Xerces&c986230&src/&3453&512\\
ArgoUML&6edc166&src\_new/&5559&1230 \\
Jedit&e343491& ./&1181&423\\ \hline
\end{tabular}
}
\end{table}

As explained in Section~\ref{Subsection: oracle}, the choice of these systems has been motivated by the need of manually-detected occurrences of God Class. However, they have been used in prior studies for a similar purpose. Consequently they belong to various domains and as shown in Table~\ref{Table: systems-hand}, they have different sizes and history lengths which comfort us for the generalization of our findings.

\subsection{Evaluation Metrics}
To compare the performances achieved by different approaches on the studied systems, we consider each approach as a binary classifier able to perform a boolean prediction on each entity of the system. Thus, we evaluate their performances using the following confusion matrix:

\begin{table}[htb]
\caption{Confusion Matrix for Anti-patterns Detection}
\label{Table: confusion matrix}
\begin{center}
\begin{tabular} {c  c | c  c | c }
& & \multicolumn{2}{ c |}{\textbf{\textit{predicted}}} & \multirow{2}{*}{\rotatebox[origin=c]{90}{\textbf{\textit{total}}}}\\ 
& & $1$ & $0$ & \\ \hline
\multirow{2}{*}{\rotatebox[origin=c]{90}{\textbf{\textit{true}}}}& $1$ & $TP$ & $FN$ & $n_{pos}$ \\
& $0$ & $FP$ & $TN$ & $n_{neg}$\\ \hline
\multicolumn{2}{ c |}{\textbf{\textit{total}}}& $m_{pos}$& $m_{neg}$ & $n$
\end{tabular}
\end{center}
\end{table}

With ($TP$) the number of true positives; ($FN$) the number of false negatives (or misses); ($FP$) the number of false positives and ($TN$) the number of true negatives. Also, we note $m_{pos}$ and $m_{neg}$ respectively the number of entities positively and negatively labeled by a classifier; $n_{pos}$ and $n_{neg}$ respectively the number of affected and healthy entities in a system ; and $n$ the total number of entities in the system. Then, based on this matrix, we compute the widely adopted \textit{precision} and \textit{recall} metrics:

\begin{center}
\begin{minipage}{.5\linewidth}
\begin{equation}
  precision = \frac{TP}{TP + FP}
\end{equation}
\end{minipage}%
\begin{minipage}{.5\linewidth}
\begin{equation}
  recall = \frac{TP}{TP + FN}
\end{equation}
\end{minipage}
\end{center}

Although these two metrics express relevant information about the classification performances, our evaluation of the classifiers investigated in this work should not rely solely on them. Indeed, commonly used evaluation metrics such as precision, recall or F-measure can be biased due to the fact that the positive and negative classes are of different sizes~\cite{powers2011evaluation}. For this reason and to obtain a single aggregated evaluation metric, we also compute the Matthews Correlation Coefficient (MCC)~\cite{matthews1975comparison}, which has been shown to limit the bias mentioned above and provide a more accurate description of the confusion matrix~\cite{powers2011evaluation}:   
\begin{equation}
\label{mcc}
MCC = \frac{TP \times n - n_{pos} \times m_{pos}}{\sqrt{n_{pos} \times m_{pos} \times n_{neg} \times m_{neg}}}
\end{equation}

\subsection{Training}
\label{subsection: training}
Here, we discuss the considerations adopted to train neural-networks on the task of anti-patterns detection. We consider training a multi-layer feed-forward neural-network to assign a boolean value to the entities of a system. First, the training set $\mathcal{D} = \{(\textbf{x}_{i}, y_{i})\}_{i=1}^{n}$ contains the instances and labels associated to each entity of the training systems. With $\textbf{x}_{i}$ the input vector (i.e., instance) corresponding to the $i^{th}$ entity, $y_{i} \in \{0, 1\}$ the true label for this entity and $n$ the size (i.e., number of entities) of the training set. Second, we refer to the set of weights of the network as $\bm{\theta} = \{\textbf{w}_{l}\}_{l=1}^{L}$, with $\textbf{w}_{l}$ being the weight matrix of the $l^{th}$ layer and $L$ the number of layers in the network. Finally, we refer to the output of the neural network corresponding to the positive label, i.e., the predicted probability that an entity is affected as:

\begin{equation}
   P_{\bm{\theta}}(1|\textbf{x}_{i}) = g(\textbf{x}_{i}.\bm{\theta}) 
\end{equation}

With $g$ the sigmoid function:

\begin{equation}
    g(x) = \frac{1}{1 + e^{-x}}
\end{equation}

\subsubsection{Custom Loss Function}
\label{section: loss}
Software systems are usually affected by a small proportion of anti-patterns ($<1\%$) \cite{palomba2018diffuseness}. As a consequence, the distribution of labels within a dataset containing software system entities is highly imbalanced. Such imbalanced dataset compromises the performances of models optimized using conventional loss functions \cite{he2008learning}. Indeed, the conventional \textit{binary\_cross\_entropy} loss function maximizes the expected accuracy on a given dataset ,i.e., the proportion of instances correctly labeled. In the context of anti-patterns, the use of this loss function lead to useless models that assign the majority label to all input instances, thus maximizing the overall accuracy ($>99\%$) during training. To overcome this issue, we must define a loss function that reflects our training objective (i.e., finding the set of weights $\bm{\theta^{*}}$ that maximizes the MCC achieved on the training set) which can be expressed as:

\begin{equation}
\bm{\theta^{*}} = \argmax_{\bm{\theta}} MCC(\bm{\theta}, \mathcal{D})
\end{equation}

Which can be addressed through gradient descent by considering a loss: $L = - MCC$. However, to compute the gradient of the loss, we need it to be a continuous and differentiable function of the weights $\bm{\theta}$. As defined in Equation~\ref{mcc}, the MCC does not meet this criterion, which prevents its direct use to define our loss function. Indeed, computing the number of true positives ($TP$) and positives ($m_{pos}$) (cf. Table~\ref{Table: confusion matrix}) requires counting elements from the output probability of the model, which necessarily involves discontinuous operators:
\begin{equation}
TP(\bm{\theta}, \mathcal{D}) = \sum_{\substack{i=1 \\ y_{i} = +1}}^{n} \delta(P_{\bm{\theta}}(1|\textbf{x}_{i}) > 0.5)
\end{equation}
\begin{equation}
m_{pos}(\bm{\theta}, \mathcal{D}) = \sum_{i=1}^{n} \delta(P_{\bm{\theta}}(1|\textbf{x}_{i}) > 0.5)
\end{equation}

With:

\begin{equation}
\label{kronecker}
    \delta(x) = \bigg\{\begin{tabular}{l}
1 if $x$=True\\
0 if $x$=False
\end{tabular}
\end{equation}

To define a continuous and differentiable approximation of the MCC, we proceed similarly to the approach proposed by Jansche~\cite{jansche2005maximum} to maximize the expected F-measure through logistic regression, which relies on the following limit:

\begin{equation}
\label{equation: approx}
\delta(P_{\bm{\theta}}(1|\textbf{x}_{i}) > 0.5) = \lim_{\gamma \to + \infty} \text{g}(\gamma\textbf{x}_{i}.\bm{\theta})
\end{equation}

Thus, to calculate the loss of our model, we compute the MCC as expressed in Equation~\ref{mcc} using the identity $m_{neg} = n - m_{pos}$, as well as the following approximations: 

\begin{equation}
TP(\bm{\theta}, \mathcal{D}) \approx \sum_{\substack{i=1 \\ y_{i} = +1}}^{n} g(\gamma \textbf{x}_{i}.\bm{\theta}) \quad with \quad \gamma > 0
\end{equation}
\begin{equation}
m_{pos}(\bm{\theta}, \mathcal{D}) \approx \sum_{i=1}^{n} g(\gamma \textbf{x}_{i}.\bm{\theta}) \quad with \quad \gamma > 0
\end{equation}

Note that the value of the hyper-parameter $\gamma$ will be adjusted during the tuning of our model.

\subsubsection{Regularization}
\label{subsection: regularization}
Regularization is a way to prevent a statistical model to learn specific and irrelevant features of its training data, which is known as ``overfitting''. To help our model to generalize to new examples and prevent overfitting, we use the widely adopted $L_{2}$ regularization technique.

$L_{2}$ regularization consists in adding a term to the loss function to encourage the weights to be small \cite{witten2016data}. This term is proportional to the sum of the Euclidean norm of the weight matrices, i.e., $\|\textbf{w}\|_{2}=\sqrt{\textbf{w}^{\top}\textbf{w}}$, also called $L_{2}$-norm. Thus, the $L_{2}$ regularization term added to the loss function can be expressed as:

\begin{equation}
L_{2} = \lambda \sum_{l=1}^{L+1} \|\textbf{w}_{l}\|_{2}
\end{equation}

With $\lambda \in \mathbb{R}$ an hyper-parameter adjusted during cross-validation.

\subsubsection{Boosting}
\label{boosting}
Before starting to train a neural-network on a given dataset, the values of each weight of the network must me set to an initial value, which is called \textit{weight initialization}. In practice, the weights are often initialized randomly following a Gaussian distribution whose parameters (i.e., mean and standard deviation) depend on the network's shape~\cite{glorot2010understanding}. As a consequence, two neural networks trained identically may achieve different performances due to the randomness of their initialization. To limit this phenomenon, several networks are trained separately and the final prediction is computed from the output of each classifier, which is commonly referred to as \textit{boosting} or \textit{ensemble learning}. Beside stabilizing the output of the model, such technique has been shown to improve the overall performances and even to lead to better results than those of each independent classifier~\cite{dietterich2000ensemble}. In the context of this study, we used the widely adopted Bayesian averaging heuristic to compute the ensemble prediction of the model from the output of several independently trained networks (10 in our study). This heuristic simply consists in taking the average of the probabilities outputed by the different classifiers. 

%% file: 05Study.tex
\section{Evaluation of the Detection Performances}
\label{section: evaluation performances}
In this section, we address the evaluation of \NAME{} in detecting the two anti-patterns considered in this study. We answer the two following research questions:

\begin{itemize}
\item \textbf{RQ1:} \textit{\RQone{}}
\item \textbf{RQ2:} \textit{\RQtwo{}}
\end{itemize}

\subsection{Study Design}
\label{subsection: study1 design}
The goal of this study is to evaluate the proposed ensemble method on both God Class and Feature Envy. We compare \NAME{} to the standalone detection tools aggregated through our approach as well as to other ensemble methods. To assess the performances of the respective approaches on each subject system, we rely on a leave-one-out cross validation. Consequently, for each of the eight systems presented in Table~\ref{Table: systems-hand}, we leave this system out for testing while keeping the other seven for hyper-parameters tuning and training. 

To run the standalone tools on the evaluation systems (\textbf{RQ1}), we used their publicly-available implementations whenever possible and replicated the approaches for which no implementation was available. Thus, we ran DECOR using the Ptidej API\footnote{https://github.com/ptidejteam/v5.2/} and JDeodorant using its Eclipse plug-in\footnote{https://marketplace.eclipse.org/content/jdeodorant/}. We implemented the detection rules for HIST as described in the original paper \cite{PalombaBPOLP13}. InCode Eclipse plug-in is no longer available and we re-implemented its detection rule as described in the original book \cite{lanza2007object} to retrieve also information about the envied class, as explained in Section~\ref{Paragragh: metrics feature envy}.

To answer \textbf{RQ2}, we choose to compare \NAME{} to the voting technique and the method ASCI proposed by Di Nucci et al.~\cite{di2017dynamic}. We selected these two techniques because, to the best of our knowledge, other ensemble methods are specific to machine-learning classifiers and cannot be adapted to our problem. Furthermore, these two techniques have been shown to achieve state-of-the-art results in the context of bug prediction \cite{wang2011software, liu2010evolutionary, di2017dynamic}. However, one must note that these methods are originally used to aggregate the output of several pre-trained machine learning classifiers while in this study, we use them to combine anti-patterns detection tools. The voting technique predicts an entity as affected if it has been detected by at least $k$ classifiers. We call $k$ the policy of the vote. This parameter has been tuned before conducting our experiments, contrary to the \textit{Validation and Voting} method proposed by Liu et al.~\cite{liu2010evolutionary}, which uses a majority voting policy ($k=2$ in our case). The ASCI method uses a decision tree algorithm to predict the best classifier for each entity given its characteristics. Then, an entity is declared as affected according to the classifier selected by ASCI for this entity. Regarding the input of the decision tree, i.e., the characteristics of the entities, we used the same metrics used for \NAME{} and presented in Section~\ref{subsection: input}. Also, similarly to a neural-network, a decision tree is subject to variations in its prediction from one training to another. Hence, to compute the performances of ASCI, we use the same boosting technique used for \NAME{} and explained in Section~\ref{boosting}. 

\subsection{Hyper-parameters Tuning}

\begin{table}
\caption{Hyper-parameters Tuning} 
\label{Table: hyper-parameters tuning}
\centering
\resizebox{9.2cm}{!}{
\begin{threeparttable}
\begin{tabular} { l l l }
\hline
Model & Hyper-parameter & Range \\ \hline
\multirow{5}{*}{\NAME{}} & $\eta$ & $10^{-[0.0; 2.5]}$\\
& $\lambda$ &  $10^{-[0.0; 2.5]}$\\
& $\gamma$ & $[1; 10]$\\
& $num\_hidden\_layers$ & $[1; 3]$ \\
& $hidden\_layers\_sizes$ & $[4; 100]$ then $[4; s]$\\ \hline
\multirow{4}{*}{ASCI} & $max\_features$ & $\{sqrt, log2, None\}$\\
& $max\_depth$ & $10\times[1; 10]$\\
& $min\_samples\_leaf$ & $[1; 5]$\\
& $min\_samples\_split$ & $10^{-[1.0; 4.0]}$\\ \hline
Vote & $k$ & $[1;3]$ \\ \hline
HIST (FE) & $\alpha$ & From 100\% to 300\% by 5\% \\ \hline
HIST (GC) & $\beta$ &  From 0\% to 20\% by 0.5\%\\ \hline
\multirow{3}{*}{InCode} & $T_{ATFD}$ & $[1; 5]$\\
& $T_{LAA}$ & $[1; 5]$\\
& $T_{FDP}$ & $[1; 5]$ \\ \hline

\end{tabular}
\begin{tablenotes}
\small
\item \textit{With $s$ the size of the previous dense layer.}
\end{tablenotes}
\end{threeparttable}
}
\end{table}

Before computing the performances achieved by each approach on a given system, we first calibrate its hyper-parameters whenever necessary. As explained in the previous subsection, this calibration is performed using instances of the seven remaining systems. The number of hyper-parameters to be tuned for \NAME{} and ASCI makes it impossible to perform an exhaustive search, thus we relied on a random search (200 random generations). This technique has shown to be more efficient than grid search on similar multi-dimensional optimization problems \cite{bergstra2012random}.
For these two approaches, we evaluate the performances achieved with each hyper-parameters' combination by carrying out a leave-one-out cross-validation. At each iteration, we thus keep instances of six systems for training (100 epochs) and use the remaining one as a validation set. We finally keep the hyper-parameters that led to the best overall MCC on the validation sets.

For \NAME{}, we tuned five hyper-parameters: the learning rate ($\eta$), the $L_{2}$ regularization weight ($\lambda$), the parameter $\gamma$ of our loss function (cf. Section~\ref{section: loss}), the number of hidden layers, and the number of neurons per hidden layer. Finally, the two models used for experiments (i.e., one model per anti-pattern) have been trained during 120 epochs with an exponential learning rate decay of 0.5 at the $100^{th}$ epoch. For ASCI, we tuned four hyper-parameters: the number of features to consider for a split ($max\_features$), the maximum depth of the decision tree ($max\_depth$), the minimum number of samples required to be at a leaf node ($min\_samples\_leaf$) and the minimum proportion of samples required to split an internal node ($min\_samples\_split$). As explained in Section~\ref{subsection: study1 design}, for the voting technique, we tuned one single parameter: the policy of the vote ($k$).

Although we followed rigorously the guidelines given by the authors of HIST and InCode when reimplementing these tools, some differences may remain between our respective implementations. Such differences could affect the optimal values of their parameters. Thus, we performed an additional parameter tuning for these tools. For HIST we tuned the parameters $\alpha$ and $\beta$ respectively related to Feature Envy and God Class detection (cf. Section~\ref{subsection: input}), while for InCode, we tuned the thresholds related to the metrics ATFD, LAA and FDP ($T_{ATFD}$, $T_{LAA}$ and $T_{FDP}$). Table~\ref{Table: hyper-parameters tuning} reports for each approach investigated, the hyper-parameters that have been tuned, as well as the ranges of values experimented.

\subsection{Analysis of the Results}
Table~\ref{Table: performances god class} reports the results of our experiments for God Class while Table~\ref{Table: performances feature envy} reports our results for Feature Envy. Our results report the performances on the eight subject systems, in terms of precision, recall, and MCC, achieved by: (1) the three detection tools used for aggregation; (2) the two competing ensemble methods: Vote and ASCI; and, (3) \NAME{}. Note that a "--" is indicated in the cells where it was not possible to compute a metric, e.g., when an approach did not detect any occurrence. To resume our results, Table~\ref{Table: overall performances} reports the overall performances achieved by each approach for God Class and Feature Envy. The overall performances are computed by merging the instances of the eight subject systems, as if they belonged to a single one.

\begin{table}
\caption{Overall Performances}
\label{Table: overall performances}
\begin{adjustbox}{width=\linewidth}
\begin{tabular}{|c|c|c|c|c|c|c|}
\hline
\multirow{2}{*}{Approaches}& 
\multicolumn{3}{c|}{
	God Class
} 
&\multicolumn{3}{c|}{
	Feature Envy
}\bigstrut [t] \\ 
\cline{2-7}
&\textit{Precision}&\textit{Recall}&\textit{  MCC  }
&\textit{Precision}&\textit{Recall}&\textit{  MCC  }
\bigstrut [t]\\
\hline
Rule Card &28\%&47\%&35\%&45\%&54\%&49\% \bigstrut \\ \hline
HIST &53\%&14\%&26\%&2\%&8\%&2\% \bigstrut \\ \hline
JDeodorant &8\%&47\%&16\%&38\%&47\%&42\% \bigstrut \\ \hline
Vote &34\%&32\%&32\%&12\%&41\%&22\% \bigstrut \\ \hline
ASCI &27\%&26\%&25\%&48\%&54\%&51\% \bigstrut \\ \hline
\textbf{\NAME{}} &45\%&51\%&\textbf{46\%}&40\%&80\%&\textbf{56\%} \bigstrut \\ \hline
\end{tabular}
\end{adjustbox}
\end{table}

\begin{table*}
\caption{Performances per System for God Class}
\label{Table: performances god class}
\begin{adjustbox}{width=\textwidth}
\begin{tabular}{|c|c|c|c|c|c|c|c|c|c|c|c|c|}
\hline
\multirow{2}{*}{Approaches}& 
\multicolumn{3}{c|}{
	Android Telephony
} 
&\multicolumn{3}{c|}{
	Android Support
}
&\multicolumn{3}{c|}{
	Apache Ant
}
&\multicolumn{3}{c|}{
	Apache Lucene 
}\bigstrut [t] \\ 
\cline{2-13}
&\textit{Precision}&\textit{Recall}&\textit{  MCC  }
&\textit{Precision}&\textit{Recall}&\textit{  MCC  }
&\textit{Precision}&\textit{Recall}&\textit{  MCC  }
&\textit{Precision}&\textit{Recall}&\textit{  MCC  } \bigstrut [t]\\
\hline
DECOR &17\%&9\%&8\%&--&0\%&--&14\%&43\%&24\%&100\%&25\%&50\% \bigstrut \\ \hline
HIST &75\%&55\%&62\%&100\%&50\%&70\%&--&0\%&--&--&0\%&-- \bigstrut \\ \hline
JDeodorant &17\%&36\%&18\%&33\%&25\%&26\%&3\%&57\%&11\%&7\%&25\%&9\% \bigstrut \\ \hline
Vote &75\%&27\%&44\%&100\%&25\%&49\%&13\%&29\%&18\%&100\%&25\%&50\% \bigstrut \\ \hline
ASCI &17\%&9\%&8\%&--&0\%&--&27\%&43\%&33\%&100\%&25\%&50\% \bigstrut \\ \hline
\textbf{\NAME{}} &0\%&0\%&-4\%&100\%&50\%&70\%&36\%&71\%&50\%&67\%&50\%&57\% \bigstrut \\ \hline
\end{tabular}
\end{adjustbox}

\bigskip

\begin{adjustbox}{width=\textwidth}
\begin{tabular}{|c|c|c|c|c|c|c|c|c|c|c|c|c|}
\hline
\multirow{2}{*}{Approaches}& 
\multicolumn{3}{c|}{
	Apache Tomcat
} 
&\multicolumn{3}{c|}{
	Apache Xerces
}
&\multicolumn{3}{c|}{
	ArgoUML
}
&\multicolumn{3}{c|}{
	Jedit 
}\bigstrut [t] \\ 
\cline{2-13}
&\textit{Precision}&\textit{Recall}&\textit{  MCC  }
&\textit{Precision}&\textit{Recall}&\textit{  MCC  }
&\textit{Precision}&\textit{Recall}&\textit{  MCC  }
&\textit{Precision}&\textit{Recall}&\textit{  MCC  } \bigstrut [t]\\
\hline
DECOR &67\%&40\%&52\%&54\%&100\%&72\%&21\%&41\%&27\%&17\%&60\%&30\% \bigstrut \\ \hline
HIST &--&0\%&--&--&0\%&--&--&0\%&--&22\%&40\%&29\% \bigstrut \\ \hline
JDeodorant &2\%&60\%&10\%&18\%&47\%&26\%&18\%&50\%&28\%&5\%&60\%&15\% \bigstrut \\ \hline
Vote &100\%&20\%&45\%&54\%&47\%&49\%&35\%&27\%&30\%&13\%&40\%&22\% \bigstrut \\ \hline
ASCI &100\%&20\%&45\%&100\%&20\%&44\%&20\%&32\%&24\%&23\%&60\%&36\% \bigstrut \\ \hline
\textbf{\NAME{}} &26\%&100\%&51\%&63\%&67\%&64\%&53\%&41\%&46\%&44\%&80\%&59\% \bigstrut \\ \hline
\end{tabular}
\end{adjustbox}

\vspace{2cm}

\caption{Performances per System for Feature Envy}
\label{Table: performances feature envy}
\begin{adjustbox}{width=\textwidth}
\begin{tabular}{|c|c|c|c|c|c|c|c|c|c|c|c|c|}
\hline
\multirow{2}{*}{Approaches}& 
\multicolumn{3}{c|}{
	Android Telephony
} 
&\multicolumn{3}{c|}{
	Android Support
}
&\multicolumn{3}{c|}{
	Apache Ant
}
&\multicolumn{3}{c|}{
	Apache Lucene 
}\bigstrut [t] \\ 
\cline{2-13}
&\textit{Precision}&\textit{Recall}&\textit{  MCC  }
&\textit{Precision}&\textit{Recall}&\textit{  MCC  }
&\textit{Precision}&\textit{Recall}&\textit{  MCC  }
&\textit{Precision}&\textit{Recall}&\textit{  MCC  } \bigstrut [t]\\
\hline
InCode &39\%&75\%&54\%&100\%&50\%&71\%&67\%&46\%&55\%&100\%&67\%&82\% \bigstrut \\ \hline
HIST &0\%&0\%&-1\%&0\%&0\%&-1\%&1\%&5\%&1\%&0&0\%&-1\% \bigstrut \\ \hline
JDeodorant &67\%&33\%&47\%&100\%&50\%&71\%&43\%&59\%&50\%&0\%&0\%&-1\% \bigstrut \\ \hline
Vote &100\%&8\%&29\%&20\%&100\%&44\%&67\%&9\%&25\%&8\%&67\%&23\% \bigstrut \\ \hline
ASCI &44\%&100\%&66\%&100\%&50\%&71\%&60\%&41\%&49\%&67\%&67\%&67\% \bigstrut \\ \hline
\textbf{\NAME{}} &50\%&83\%&64\%&67\%&100\%&82\%&37\%&86\%&56\%&15\%&67\%&32\% \bigstrut \\ \hline
\end{tabular}
\end{adjustbox}

\bigskip

\begin{adjustbox}{width=\textwidth}
\begin{tabular}{|c|c|c|c|c|c|c|c|c|c|c|c|c|}
\hline
\multirow{2}{*}{Approaches}& 
\multicolumn{3}{c|}{
	Apache Tomcat
} 
&\multicolumn{3}{c|}{
	Apache Xerces
}
&\multicolumn{3}{c|}{
	ArgoUML
}
&\multicolumn{3}{c|}{
	Jedit 
}\bigstrut [t] \\ 
\cline{2-13}
&\textit{Precision}&\textit{Recall}&\textit{  MCC  }
&\textit{Precision}&\textit{Recall}&\textit{  MCC  }
&\textit{Precision}&\textit{Recall}&\textit{  MCC  }
&\textit{Precision}&\textit{Recall}&\textit{  MCC  } \bigstrut [t]\\
\hline
InCode &57\%&52\%&54\%&31\%&78\%&49\%&67\%&18\%&35\%&50\%&53\%&51\% \bigstrut \\ \hline
HIST &5\%&15\%&8\%&1\%&6\%&1\%&1\%&5\%&1\%&1\%&6\%&0\% \bigstrut \\ \hline
JDeodorant &31\%&50\%&39\%&43\%&50\%&46\%&46\%&23\%&32\%&44\%&65\%&53\% \bigstrut \\ \hline
Vote &73\%&17\%&35\%&12\%&100\%&34\%&--&0\%&--&9\%&100\%&28\% \bigstrut \\ \hline
ASCI &60\%&50\%&54\%&36\%&78\%&52\%&80\%&18\%&38\%&50\%&53\%&51\% \bigstrut \\ \hline
\textbf{\NAME{}} &39\%&81\%&56\%&36\%&75\%&51\%&47\%&68\%&56\%&48\%&88\%&65\% \bigstrut \\ \hline
\end{tabular}
\end{adjustbox}
\end{table*}

\subsubsection{\RQone{}}
For God Class detection, \NAME{} shows a precision of $45\%$, a recall of $51\%$ and a Matthew's Correlation Coefficient of $0.46$ on overall over the subject systems. Thus, the proposed ensemble method clearly outperforms the three approaches used for aggregation. Specifically, the overall MCC improves by $45\%$ in comparison to the tool that performed the best (DECOR with $0.35$). Considering the performances achieved on each system, we can see that \NAME{} achieves poor performances on the first system (Android Telephony) due to our model over-fitting its training data. However on the remaining seven systems the proposed method shows a MCC ranging between $0.46$ and $0.70$, which confirms that \NAME{} performs well independently of the systems characteristics. On the contrary, the competing tools show wider ranges of MCC and achieve poor performances on several systems. However, the low performances reported for JDeodorant (especially precision) can be due to this tool relying on a different definition of God Class than others. Indeed, affected entities are detected only if opportunities to split them are identified.

For Feature Envy detection, \NAME{} achieves on overall a precision of $40\%$ and a recall of $80\%$ leading to an MCC of $0.56$. We observe better performances in terms of MCC achieved by the static code analysis tools ($0.49$ by InCode and $0.42$ by JDeodorant) than for God Class detection and low results for HIST. These results show that \NAME{} outperforms the standalone tools when detecting Feature Envy with an overall MCC $14\%$ higher than that of the tool that performed the best (InCode). Similarly to God Class, \NAME{} shows acceptable performances on every systems with an MCC ranging between $0.32$ and $0.82$. It is important to highlight that when replicating HIST's rule for Feature Envy detection, we used a different component\footnote{http://www.incava.org/projects/diffj} to extract changes at method level than that of the original approach because the original component is supposedly unavailable because of its license. We are aware that such difference could affect the reported performances.

\Answer{\NAME{} significantly outperforms the standalone detection tools in detecting God Class and Feature Envy. Furthermore, our results indicate that \NAME{} performs well independently of the systems characteristics.}

\subsubsection{\RQtwo{}}
We report the results of our study aiming to compare the performances of \NAME{} with those of the two competing ensemble methods considering God Class and Feature Envy independently in turn. For God Class, we can see that our method outperforms on overall both the voting technique and ASCI in terms of precision, recall and MCC. Indeed, for this anti-pattern none of the competing ensemble method seems to improve the performances of the aggregated tools. However when looking at the performances achieved on each system, we observe that contrary to ASCI, the voting technique has the advantage of showing an acceptable MCC for every systems. However, we must notice that in our experiments we chose to feed the decision tree used by ASCI with the same metrics used to feed the MLP on which rely our approach \NAME{}. We cannot exclude that another version of ASCI based on different software metrics may have led to better results.

For Feature Envy, we can see that in term of precision, ASCI outperforms our method with $48\%$ against $40\%$ for \NAME{}. However, \NAME{} achieves a recall significantly higher than its competitors with a value of $80\%$ against $54\%$ and $41\%$. Finally, regarding the MCC, we can see that although ASCI is able to improve the performances of the aggregated tools, its is our method, \NAME{} that shows the biggest improvement with an MCC of $0.56$

\Answer{\NAME{} significantly outperforms both the voting technique and ASCI in terms of recall and MCC. Furthermore, our results indicate that none of the competing method is able to improve the detection for both anti-patterns.}

%% file: 06Threats.tex
\section{Threats to Validity}
\label{section: threats}

In this section, we discuss the threats that could affect the validity of our study.

\paragraph{Construct Validity} Threats to construct validity concern the relation between theory and observation. In our context, this could refer to the reliability of the oracle used to train and evaluate the different approaches investigated in this work. Instances of God Class extracted from HIST and DECOR replication packages have been filtered before being incorporated in our oracle. Furthermore, both papers have been awarded by the community, which confirms the quality of the processes conducted to produce these instances. For Feature Envy, we followed a strict blind procedure where each instance has been investigated by three different persons. However, we can not exclude the possibility of some missed occurrences or false positives. Another threat is related to the replication of some of the competitive approaches. We followed rigorously the guidelines provided by the respective authors, and performed an additional parameter tuning for each approach whenever necessary. However, some differences may remain between our respective implementations.

\paragraph{Internal Validity} Threats to internal validity concern all the factors that could have impacted our results. In our context, this could refer to the training procedure presented in Section~\ref{subsection: training}. Even though we compared the proposed procedure with conventional techniques while performing preliminary experiments, we did not report the results of our comparisons. Hence, a comparative study of the proposed procedure with conventional optimization approaches would be desirable. Note that these techniques are in fact part of the approach we propose and are necessary to train models on real imbalanced datasets. Another threat is related to choice of the machine-learning based classifier (MLP) used in our method. We plan to investigate the use of different machine-learning algorithms to perform aggregation. Finally, to train the competing ensemble method ASCI, we first performed an hyper-parameter tuning of this approach rigorously identical to that of \NAME{} and used the same boosting technique.

\paragraph{External Validity} Threats to external validity concern the generalizability of our findings. To reduce this threat, we experimented our method on a reasonable number of systems (8). Furthermore, the software systems used for evaluation have different domains, origins, sizes and history lengths. However, further evaluation of our approach on a larger set of systems would be desirable. Another threat could be related to the choice of the detection tools chosen to be aggregated. As explained in Section~\ref{subsection: overview} we selected these tools because they are based on different strategies, and thus, are more likely to have complementary results. However, we cannot assert that using \NAME{} to aggregate other detection tools would led to similar results.

%% file: 07Conclusion.tex
\section{Conclusion and Future Work}
\label{section: conclusion}

We proposed \NAME{}, a machine-learning based ensemble method to aggregate various anti-pattern detection approaches on the basis of their internal detection rule. The goal is to create an improved classifier with respect to the standalone approaches. Our method consists in identifying the core-metrics of each approach to be aggregated and then use these metrics to feed a neural-network classifier. To train and evaluate our model, we built an oracle containing the occurrences of God Class and Feature Envy in eight open-source systems. To address the poor performances commonly reported by neural-networks on imbalanced datasets such as our oracle, we also designed a training procedure allowing to maximize the expected MCC. Then, we evaluated \NAME{} on this oracle and compared its performances with the so aggregated tools as well as with competing ensemble methods. Key results of our experiments indicate that:

\begin{itemize}
\item \NAME{} significantly outperforms the standalone tools aggregated through our approach in detecting God Class and Feature Envy and performs well independently of the systems characteristics.

\item \NAME{} outperforms other ensemble methods in terms of recall and MCC. Also, none of the competing ensemble method (i.e., Vote and ASCI) has succeeded to improve the detection for both anti-patterns.
\end{itemize}

Future work includes a comparative study of the different machine-learning algorithms that could be used for aggregation. We also plan to extend our approach to the detection of other anti-patterns with a greater number of detection tools.